# A Comparison between the *China Scientific and Technical Papers and Citations Database* and the *Science Citation Index* in terms of journal hierarchies and inter-journal citation relations

Ping Zhou[ab] & Loet Leydesdorff[b]

[a]Institute of Scientific and Technical Information of China.
15 Fuxing Road. Beijing 100038. P. R. China; zhoup@istic.ac.cn

[b]Amsterdam School of Communications Research (ASCoR). University of Amsterdam
Kloveniersburgwal 48. 1012 CX Amsterdam. The Netherlands;
loet@leydesdorff.net ; http://www.leydesdorff.net

**Abstract**

The journal structure in the *China Scientific and Technical Papers and Citations Database (CSTPCD)* is analysed from three perspectives: the database level, the specialty level and the institutional level (i.e., university journals versus journals issued by the Chinese Academy of Sciences). The results are compared with those for (Chinese) journals included in the *Science Citation Index*. The frequency of journal-journal citation relations in the *CSTPCD* is an order of magnitude lower than in the *SCI*. Chinese journals, especially high-quality journals, prefer to cite international journals rather than domestic ones. However, Chinese journals do not get an equivalent reception from their international counterparts. The international visibility of Chinese journals is low, but varies among fields of science. Journals of the Chinese Academy of Sciences (CAS) have a better reception in the international scientific community than university journals.

# 1. Introduction

With the continuous development of the Chinese economy, the scientific production of China is also experiencing notable growth. Take scientific publications as an example: the percentage of the world share of Chinese publications increased exponentially during the period 1993-2004 (Jin & Rousseau, 2004; Zhou & Leydesdorff, 2006). This increase advanced China's position from the $17^{th}$ in 1993 to the $5^{th}$ in 2004 (ISTIC, 1998). However, the number of citations received by Chinese publications is low. In 2004, China ranked only the $14^{th}$ on this indicator (ISTIC, 2005). Although this is a big advancement compared to the $18^{th}$ position in 2003, the performance of China in terms of publications and citations is not yet compatible.

In order to investigate the reasons for these relatively low citation rates, it would theoretically be interesting to compare Chinese articles with their Western counterparts as matched pairs in terms of their quality and the number of citations received. However, it is difficult to assess quality independent of citation rates and a critical problem is that papers can be cited for a variety of reasons. Some authors cite papers in order to place their contributions in relevant discussions, some references serve as summaries, and others are used as additional warrant of the knowledge claims (Leydesdorff & Amsterdamska, 1990). Although finding matched pairs of papers for the comparison thus may be virtually impossible, comparison at an aggregated level, such as at the level of nations or journals, is feasible.

In addition to the intrinsic quality of articles, other factors like language and the availability of the journal online may affect the visibility of a journal. No research has been done about the visibility of Chinese journals as a possible cause for the low citation rates. We use routines developed by one of us in the context of the international set of the ISI-journals (Leydesdorff & Cozzens, 1993) for studying the position and visibility of Chinese journals. The visualizations are based on using Pajek. Citation data is collected from the *China Scientific and Technical Papers and Citations Database (CSTPCD)* and the *Science Citation Index (SCI)*, respectively.

Since its first publication in 1964, the *Science Citation Index* (*SCI*) has been widely used by universities, research institutions, and individuals to evaluate research output. In 2003, 5,907 journals from various countries were included as sources of the *SCI*. To some extent, the *SCI* data can represent a country's scientific production (Sivertsen, 2003). It cannot, however, provide the full panorama of a country's scientific output, especially not when the official language is not English. Inclusion in the *SCI* has been debated in terms of national, language, and disciplinary biases. Van Leeuwen *et al.* (2001), for example, argued that the language bias of coverage can have consequences for international comparisons of national research performance.

There were 4,497 scientific journals published in mainland China in 2003 (Ren, 2005). Among these only 67 were included in the *SCI* of that year (that is, about 1.5%). Thus, it is imperative for a country like China—a large country that has five thousand years of history and a tradition of nurturing science and education—to formulate a database for the purpose of evaluating its scientific outputs. The *China Scientific and Technical Papers and Citation Database* (*CSTPCD*), a database similar to the *SCI*, was set up in 1988 with the support of the Ministry of Science and Technology. The Institute of Scientific and Technical Information of China (ISTIC) has carried out and developed the project ever since, making the *CSTPCD* widely used by research institutions, scientific management organizations, and individual scientists to measure their research output (Wu *et al.*, 2004).

When the database was first established in 1988, only 1,189 journals were included; 15 years later (2003), this number has increased to 1,576 journals. The annual news conference on the statistics of Chinese scientific publications and citations—held by the ISTIC—has been an important event in the Chinese science community. The results are published by major Chinese media, such as China Central Television Station (CCTV) and *Chinese S&T Daily*.

The *Chinese Science Citation Database (CSCD)* is another database similar to the *CSTPCD*. This database is produced by the Documentation and Information Centre of the Chinese Academy of Sciences (DICCAS), and covered 1,046 journals in 2001 (Jin & Wang, 1999). Leydesdorff and Jin (2005) used the *CSCD* to map the Chinese journal-journal citation relations.

In the current study, we use the *CSTPCD* as the data source. Among the issues which we wish to examine, are the following:

- ✧ The similarities and differences between the domestic and the international databases. Although the *SCI* and the *CSTPCD* are both widely used for research evaluation in China, and comparative studies on the two databases were done before (Liang, et al, 2001; Liang, 2003), we wish to explore this issue from the perspective of evaluating the databases using scientific journals as units of analysis;
- ✧ In order to classify journal hierarchies and layers of communication in Chinese and international journals, the aggregated journal-journal citation relations in the two databases provide information about disciplinary similarities and citation preferences among journals. Different journals have different citation impacts, and some journals are cited more frequently than others. This information can be used to classify journal hierarchies and layers of communication.
- ✧ Comparative studies at the journal level may help us to reach the above objectives.

## 2. Methods and materials

In order to visualize aggregated journal-journal citations, we use a series of previously developed routines for analysing journal-journal citation relations and the software package Pajek (available at http://vlado.fmf.uni-lj.si/pub/networks/pajek/). The aggregated journal-journal citations can be considered as a huge matrix of citing and cited journals, respectively. This matrix is asymmetrical and overwhelmingly empty. Scientific journals tend to cite one another in dense clusters that represent specialties (Leydesdorff, 2004). Some (e.g., interdisciplinary) journals cite and are cited across different fields, but the majority of the journals are embedded in a specialized publication and citation structure (Narin *et al.*, 1972). In other words, the matrix is nearly decomposable into specialty structures (Simon, 1973).

The classification of journals into their local densities has not been a *sine cura* (Doreian & Fararo, 1985; Leydesdorff, 1986; Tijssen *et al.*, 1987). Although the densities reflecting specialties are reproduced from year to year, the decomposition in each year is sensitive to the choices of the various parameters involved, such as the seed journal(s) for collecting a citation environment, the threshold levels, similarity criteria, and the clustering algorithm. In other words, the vectors of the journal distribution span a multi-dimensional space in which clouds can be distinguished, but the delineation of these clouds remains fuzzy at the edges (Bensman, 2001) and varies with the perspectives chosen by the analyst.

Leydesdorff & Cozzens (1993) developed a series of routines that generate aggregated journal-journal citation matrices on the basis of a seed journal or a set of seed journals. For this study, we modified these routines in order to differentiate between the journal environments in the citing and cited dimensions. These two environments can be very asymmetrical for the same journal found in the international database or the Chinese database. As we will demonstrate below, some journals are heavily cited domestically, but cite only internationally. The new routines generate an aggregated journal-journal citing network that includes only journals that are cited by the seed journal above a certain threshold (e.g., 1% of its total citing), while a cited network covers journals that cite the seed journal above the threshold (i.e., 1% of its total cited).

The various citation matrices are imported into SPSS for factor analysis, and read into Pajek for the visualization. The matrices were normalized using the cosine as the similarity measure (Salton & McGill, 1983). As a similarity measure, the cosine is equivalent to the Pearson correlation coefficient (Jones & Furnas, 1987), but it has advantages in the case of sparse matrices (Ahlgren *et al.*, 2003). For the purpose of the visualization, it is convenient that the cosine provides us with positive values only, while one also expects negative values in a Pearson correlation matrix. The Pearson correlation remains the analytical instrument for finding the eigenvectors of the network (e.g., by using factor analysis), while the cosine is the appropriate measure for mapping and visualizing the vector-space. The pictures included in this study only exhibit cosine values $\geq 0.2$.

Data sources originated from the 2003 aggregated journal-journal citation databases of the *China Scientific and Technical Papers and Citations Database (CSTPC)* and the corresponding *Journal Citation Report 2003* of the *Science Citation Index* (*SCI*). The results are described in three subsections: the first provides descriptive statistics about the *CSTPCD* in comparison with the *SCI*; the second subsection shows comparative results with the *SCI* in fields like general science, biology, and material science; in the third subsection we use these methods to compare the citation status of institutional journals with an origin at a Chinese university and the Chinese Academy of Sciences, respectively.

## 3. Results

*3.1*  The *CSTPCD* and the *SCI*

Table 1 contains several terms which can be derived from the two databases. Based on the original data of the *CSTPCD* or the *SCI*, we create two databases for both the *CSTPCD* and the *SCI*. The first two databases contain fields like journal names, number of citations received, and number of references. By aggregating data in the fields of the number of references and the number of citations received, respectively, one can obtain the 'total number of references' and 'total number citations received' in a database. For the 'sum of journal-journal relations', two other databases were generated with aggregated information of citations among each two journals. Based on this database, one can aggregate the total number of 'unique journal-journal relations' among journals. The word 'unique' means that if an article in journal A cites an article in journal B for one or more times, the number of citation relations is only counted as one.

Among the 1,576 journals in the *CSTPC*, 157,659 citation relations are maintained; that is, 2.3% of the 2,483,776 (= $1576^2$) possible relations. The corresponding figure is 2.8% for the *Science Citation Index* in the year 2003. These figures show that the percentage of actual journal-journal citation relations over the possible number of journal-journal relations is very low, but it is even less in the *CSTPCD* than in the *SCI* (Table 1).

| 2003 | *CSTPC* | *SCI* |
|---|---|---|
| Number of source journals processed | 1,576 | 5,907 |
| Unique journal-journal relations | 157,659 | 971,502 |
| | 2.3% | 2.8% |
| Sum of journal-journal relations | 573,543 | 17,604,594 |
| Total 'citing' | 2,233,524 | 23,953,246 |
| Total 'cited' | 570,384 | 19,497,302 |

**Table 1.** Comparison of the data in various relevant dimensions for the *CSTPCD* 2003 and the *SCI* 2003.

In the *CSTPC*, the mean of the journal-journal citation relations per journal is 364 (= 573,543 ÷ 1,576), while that for the *SCI* is 2,980 (= 17,604,594 ÷ 5,907). In other words, journal-journal citation relations are expected to occur in the *SCI* eight times (= 2980 ÷364) more than that in the *CSTPCD*. With regard to the average number of references per journal, the corresponding figures for the *CSTPCD* and the *SCI* are 1,417 (= 2,233,524 ÷ 1,756) and 4,055 (= 23,953,246 ÷ 5,907), respectively. Thus, the figure of the *SCI* is approximately three times (= 4,055 ÷ 1,417) that of the *CSTPCD*. The average number of citations per journal in the *CSTPCD* is 362, and that of the *SCI* is 3,301; the latter is nine times more frequent than the former. Thus, the citation density in the *CSTPCD* is approximately an order of magnitude lower than the density in the ISI database.

Among the 1,576 journals covered by the *CSTPCD* in 2003, 29 were published in English and the remainder (1,547 journals) were published in Chinese. However, some of the journals published in Chinese provided abstracts in English. The *Science Citation Index* (*SCI*) covered 5,907 journals in 2003, of which only 67 Chinese journals are included (1.13%). Among these 67 Chinese journals, 22 are published in Chinese and 45 are in English.

**3.2 Comparison at the level of specialties**

We selected journals in general science, material science, and the life sciences in order to compare citing and cited environments in both the international and domestic databases. Journals in general science aim to cover publications in various existing disciplines; our objective was to test whether this is the case for Chinese journals as well. According to a report of the Documentation and Information Centre of the Chinese Academy of Sciences (DICCAS, 2004), material science and mathematics are the fields in which China performs best, while the life sciences lag behind.

In general, the criteria for selecting sample journals were the following:

a. A journal is included in both the *CSTPCD* and the *SCI*. In the case of the analysis of journals in material science and the life sciences we use this criterion;
b. Some Chinese journals have both a Chinese and an English edition. The Chinese editions of this kind of journals are usually included in the *CSTPCD* since the database is mainly focused on publications in Chinese, while the *SCI* only covers the English editions of this type of journals (Ren & Rousseau, 2004). Journals under discussion in this paper which fulfil this criterion are: the *Chinese Science Bulletin, Science in China Series C, Science in China Series E,* and the *Journal of* *the University of Science and Technology Beijing*. In these cases, we use Chinese editions in the *CSTPCD* and their English editions in the *SCI* for the analysis.

*3.2.1 Journals in general science*

We selected the *Chinese Science Bulletin* (*CSB*) as the subject of study since this is considered as the most important journal in Chinese general science. According to the statistics of ISTIC, the Chinese edition of this journal ranked first in the general science class with an impact factor of 0.891 in the *CSTPCD* in 2003 (ISTIC, 2004). The journal is published in two independent editions, in Chinese and English, respectively. The English edition of *Chinese Science Bulletin* had an impact factor of 0.593 in the *SCI* in 2003 (JCR, 2003). It should be notified that the two-edition publication mechanism of some Chinese journals can cause errors in assigning citations. For example, a citation is sometimes attributed to the English edition of *Chinese Science Bulletin* although it is made to an article in the Chinese edition (Ren & Rousseau, 2002).

a. Citation environment of the *Chinese Science Bulletin* (Chinese edition) in the *CSTPC*

a.1 Citing pattern

The *Chinese Science Bulletin* (Chinese edition; we indicate this edition with *CSBC* below) had a total of 11,506 references in 2003, among which 1,605 were provided to 284 journals included in the *CSTPCD*. This means that articles in *CSBC* give only 14% of their references to journals covered by the *CSTPCD*. When the threshold is set at the convenient value of 1% (given the expectation of a Lotka distribution), there would be no other *CSTPCD* journals included in the citing environment of *CSBC* except *CSBC* itself. In other words, no other journals included in the *CSTPCD* received 1% of the *CSBC*'s total citations. Where have the rest of 9,901 (= 11,506 - 1,605) references gone? We conjecture that most of them have been given to international journals: when making citations, authors in *CSBC* favour international journals. International journals would then account for approximately 86% of the total references of *CSBC*.[1]

*CSBC* cited 125 journals only once; 49 journals were cited two times and 34 journals three times. In other words, 73% of the 284 journals were cited by the *CSBC* less than four times. Journals that obtain higher number of citation in *CSBC* were *CSBC* itself (332 times), *Science in China D* (Chinese edition; 98 times), *Quaternary Sciences* (49 times), *Acta Petrologica Sinica* (45 times), and *Science in China B* (43 times). Except the *CSBC*—a general science journal—and *Science in China B*—a chemistry journal—these journals were classified by ISTIC as belonging to the geo-sciences.

a.2 Cited pattern

700 journals included in the *CSTPCD* cited the *CSBC* in 2003, providing 3,958 citations. Among the 1,576 total journals included in the *CSTPC*, almost half (44%) had cited the *CSBC*. This means that *CSBC* has a very high visibility among Chinese scientific and technological journals. The visibility of *CSBC* matches its reputation as an important journal in general science. According to ISTIC's statistics, the journal's impact factor ranks first among journals in general science in 2003 (ISTIC, 2004).

In order to analyse which fields have a close citation relation with *CSBC*, we collected the cited environment of *CSBC* by setting the threshold as 1% (Figure 1). Each of the ten journals included in this cited environment comprised a number of citations of more than 1% of the total number of citations of *CSBC*. Among these, seven were from geology, two from general science (including *CSBC* itself), and one from geography. Thus, we may conclude that the main impact of the Chinese edition of *Chinese Science Bulletin* is also in the geo-sciences.

---

[1] It is possible that *CSBC* cited some journals that were not included in the *CSTPC*, but this number would not play a key role, since the database covers most of the Chinese journals with sufficient quality.

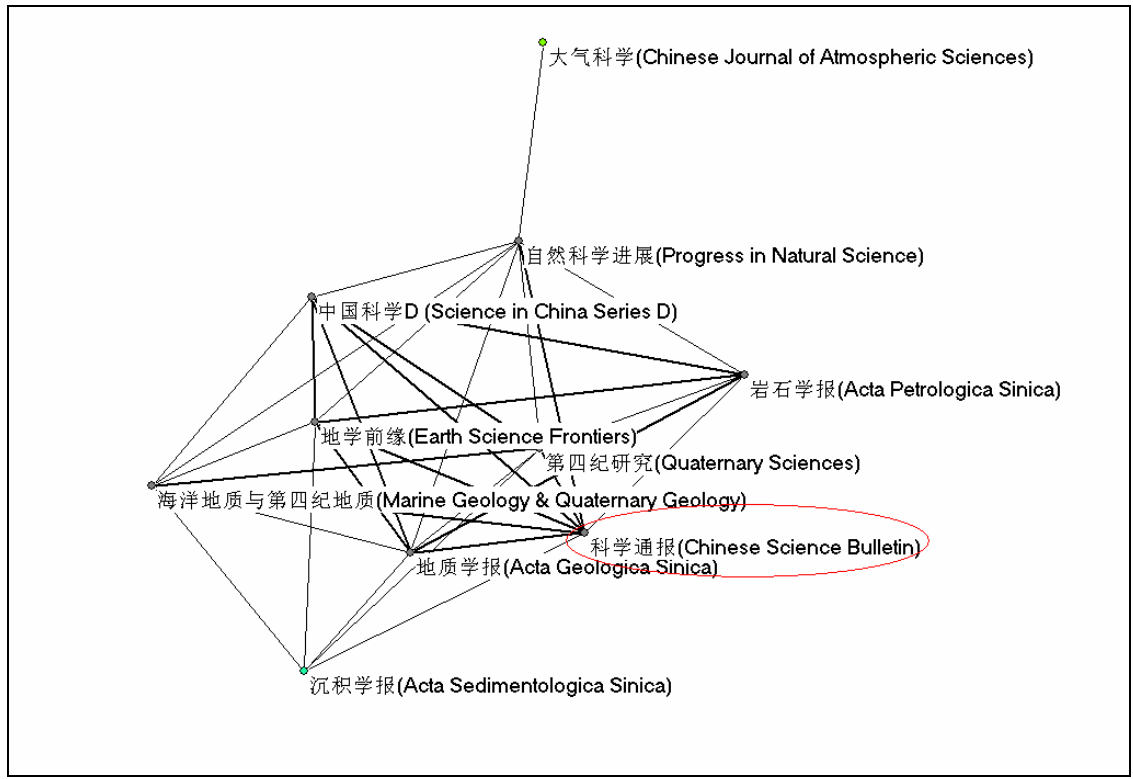

**Figure 1:** Cited environment of *Chinese Science Bulletin* (Chinese edition. *CSTPCD* 2003; threshold = 1%; cosine ≥ 0.2)

Although *CSBC* is a general science journal, its citation reception is mainly in the geosciences. The journal has a high visibility among Chinese S&T journals, but authors prefer international journals as sources for citations when publishing in *CSBC*. Does *CSBC* receive the same return from its international counterparts? We selected the English edition of *Chinese Science Bulletin* to explore this question.

b. Citation environment of the *Chinese Science Bulletin* (English edition) in the *SCI*
b.1 Citing pattern

The *Chinese Science Bulletin*—we shall denote the English version as *CSB-E* below—cited a total of 1,168 journals two or more times in 2003, and cited another 2,399 journals only once, generating 12,082 citations in total (JCR, 2003). Among these journals, 775 journals are included in the *SCI* and are cited two or more times (8,210 citations in total).

This means that at least 68% of the references by *CSB-E* are given to journals included in the *SCI*. In other words, articles in *CSB-E* tend to cite journals included in the *SCI*. Consequently, international journals have a very significant citation impact on Chinese authors who publish articles in *CSB-E*. Among the journals cited more frequently by these authors, leading and general science journals prevail (Figure 2).

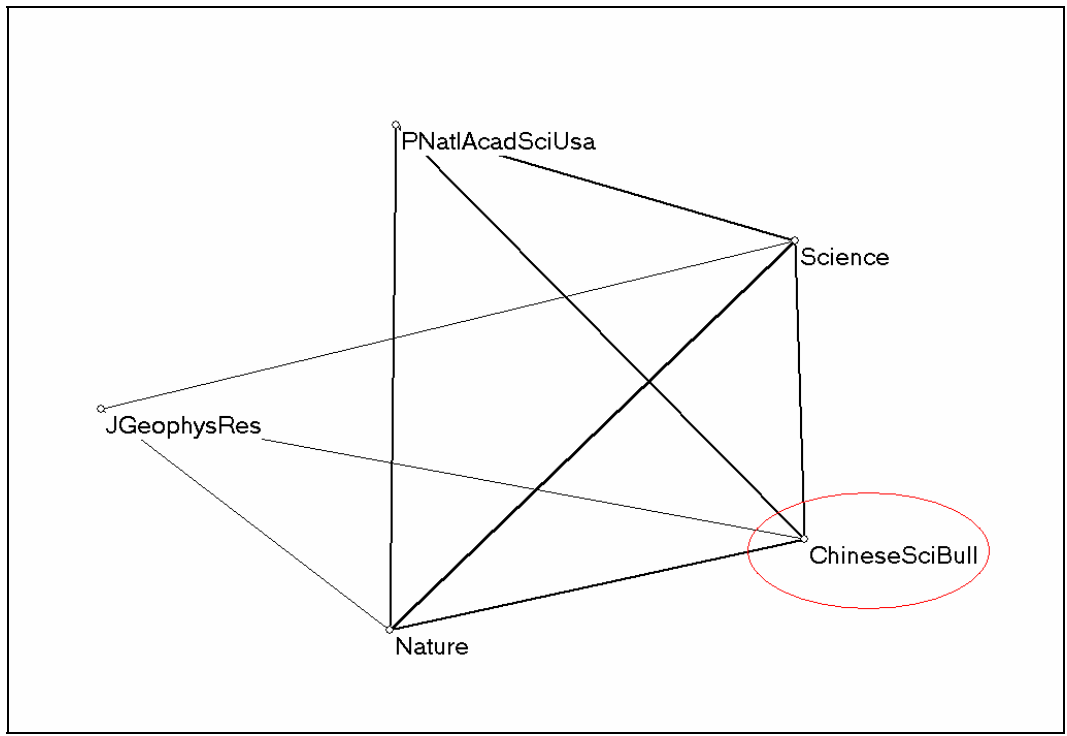

**Figure 2:** Citing network of the *Chinese Science Bulletin* (English Edition; *SCI*, 2003; threshold = 1%; cosine ≥ 0.2).

b.2 Cited pattern

*CSB-E* was cited 2,302 times in 2003, among which 1,900 citations were from 248 journals that were included in the *SCI* and provided two or more citations. The 46 Chinese journals that were included in the *SCI* and cited by *CSB-E* two or more times made a total number of citations of 1,091, with a share of 47% of the 2,302 citations. International journals account for 33% of the 2,302 citations. Among the 46 Chinese journals, 25 are published in English with a share of 33% of the 2,302 citations, and the contribution of the 21 journals in Chinese is 14%. The rest of 18% of citations were from journals that only cited the *CSB-E* once, or from journals with ambiguous information in the *SCI* database about their national origin. This means that the *CSB-E* is mostly cited by Chinese journals (Table 2).

| **Journal Name** | **Chinese journals** | **International journals** | **Other journals** |
|---|---|---|---|
| *Chinese Science Bulletin* (English edition) | 47% | 33% | 20% |
| *Journal of Inorganic Materials* (Chinese edition) | 56% | 19% | 25% |
| *Journal of Materials Science & Technology* (English edition) | 21% | 46% | 23% |
| *Science in China Series C-Life Science* (English edition) | 38% | 30% | 22% |
| *Journal of University of Science and Technology Beijing* | 64% | 11% | 25% |
| *Science in China Series E* | 38% | 22% | 40% |

Note: The title "other journals" indicates those cited the object journal only once or those with unclear information.
**Table 2:** Citation distribution of relevant Chinese journals (*SCI*, 2003).

*CSB-E* was cited by journals in various disciplines, but most of the journals citing the *CSB-E* are Chinese journals; all journals that contributed more than 1% of the total number of citations of the *CSB-E* were from China (Figure 3).

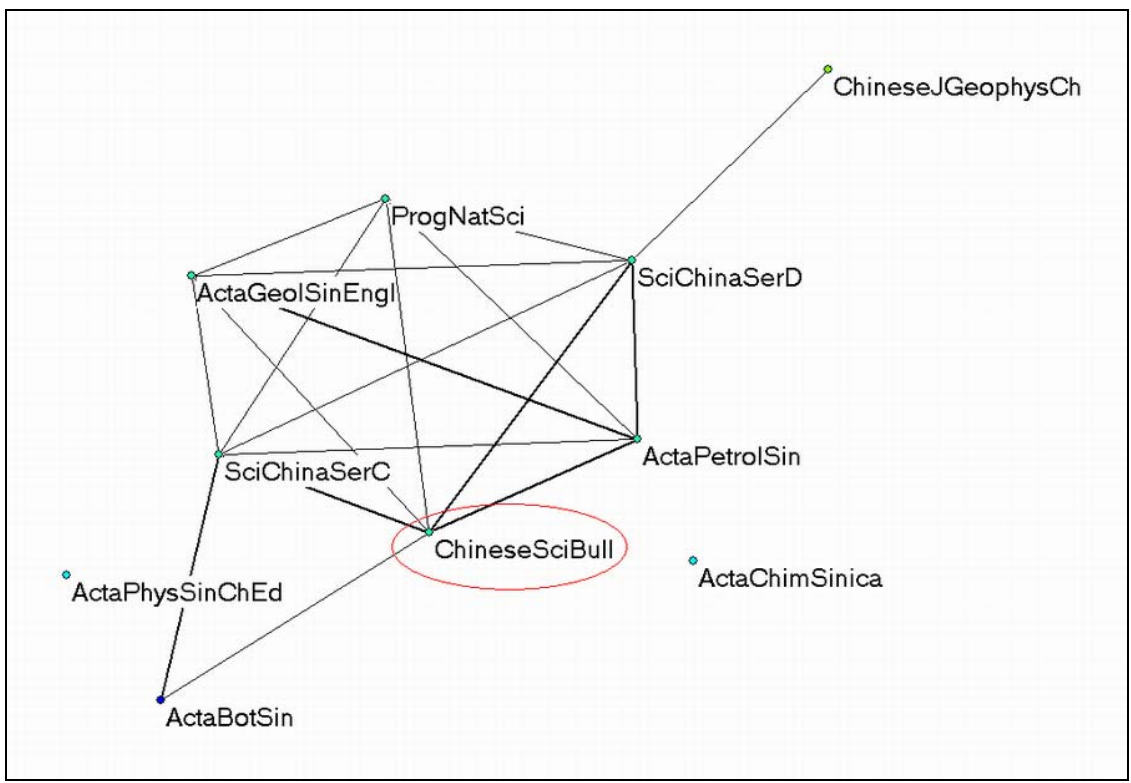

**Figure 3:** Cited pattern of the *Chinese Science Bulletin* (English Edition; *SCI*, 2003; threshold = 1%; cosine ≥ 0.2).

In summary, the *Chinese Science Bulletin* (Chinese version) is considered as an important journal in China, with the highest impact factor (0.891) among journals included in the *CSTPCD* in the category of general science in 2003. However, in terms of international visibility, Chinese journals in this field still have a long way to go. With an impact factor of 0.593, the *Chinese Science Bulletin* (*English Edition*) ranked only 18[th] among the multidisciplinary journals included in the *SCI* 2003. *Nature*'s impact factor of 30.979 is more than 52 times that of the *Chinese Science Bulletin.*

Although the two editions of the *Chinese Science Bulletin* contain the same articles, they seem to have a different disciplinary impact in terms of their citation relations. The impact of the Chinese edition is mainly focused on the geo-sciences, while the English edition behaves more like a multidisciplinary journal when evaluated in terms of its citation patterns.

*3.2.2 Journals in material science*

Here, we will examine two Chinese journals. One is the *Journal of Inorganic Materials (JIM)* published in Chinese. We chose this journal for comparison because it is covered by both the *CSTPCD* and the *SCI* in 2003. The other one is the *Journal of Materials Science & Technology*, which is published in English and is only covered by the *SCI*; the CSTPCD mainly covers journals in Chinese. We include this English-language journal in the study in order to assess whether the use of one language or the other influences a journal's visibility.

a. Citation environment of the *Journal of Inorganic Materials (JIM)* in *CSTPC*

a.1 Citing pattern

*JIM* had 3,279 references in 2003, among which 407 citations are provided to 122 journals in the *CSTPCD* (14%). As authors tend to consult research output published in relatively high-quality journals, we conjecture that the remaining 2,872 (= 3,279 – 407) references were given to international journals instead of domestic journals not covered by the *CSTPCD*. In this case, the international journals would account for 86% of the number of references of *JIM*.

The journal that was mainly cited by *JIM* (93 times out of 407) is *JIM* itself. The *Journal of the Chinese Ceramic Society* had the second highest number of references (25 times) from *JIM*. This means that *JIM* mainly cites journals in material science. Seventy journals are cited only once. When the threshold is set at 1%, no other journals in the *CSTPCD* are included in the citing environment of *JIM*, except *JIM* itself. In summary: authors publishing in *JIM* hardly cite domestic journals.

a.2 Cited pattern

Even though authors in *JIM* show little citation interest in its domestic counterparts, its visibility among Chinese journals is high. In 2003, 212 journals in the *CSTPCD* cited *JIM*, generating 896 citations. There are 20 domestic journals in the cited environment of *JIM* when the threshold is set at 1% (Figure 8). This means that the *Journal of Inorganic Materials* has high visibility in domestic material science and other relevant fields.

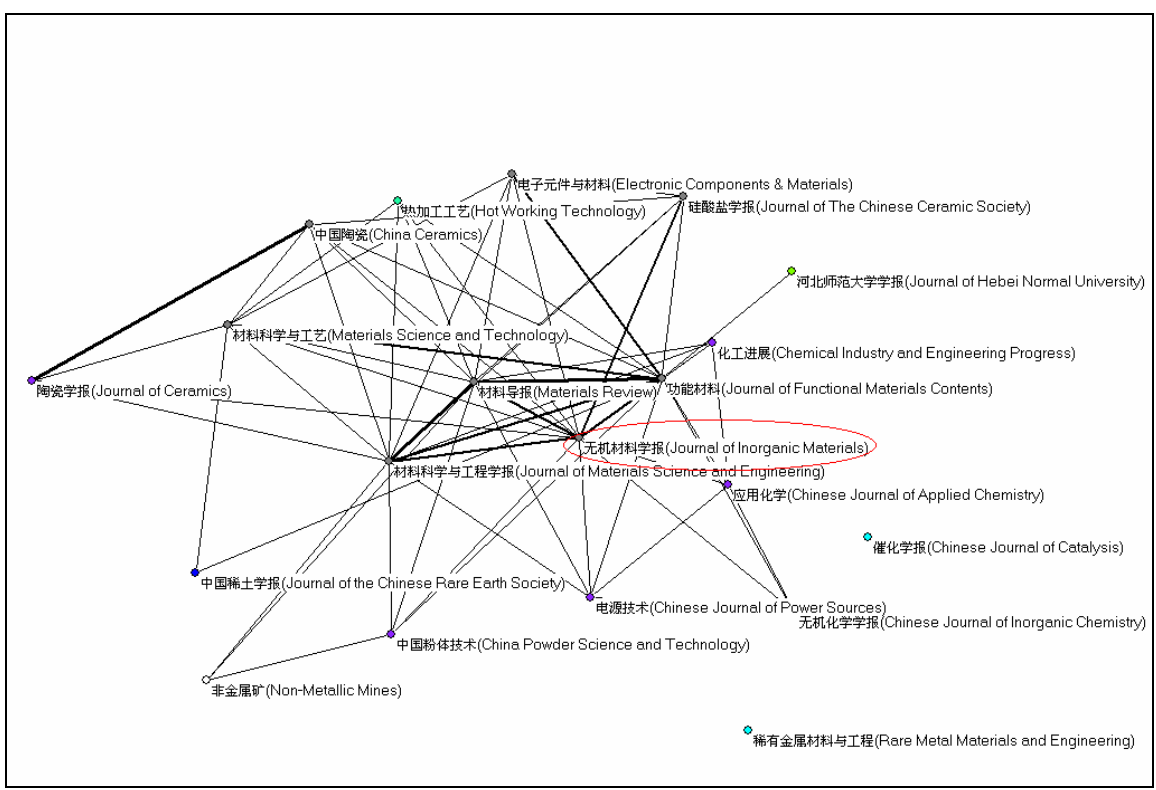

**Figure 4:** Cited environment of the *Journal of Inorganic Materials* (*CSTPCD* 2003; threshold = 1%; cosine ≥ 0.2).

b. Citation environment of the *Journal of Inorganic Materials* in the *SCI*

b.1 Citing pattern

*JIM* had 2,788 total references recorded in the *SCI* in 2003, which cited 170 journals that are covered by the *SCI* for a total of 2,249 times. (Each of the 170 journals was cited at least twice.) Figure 5 is obtained when the threshold is set at 1%. Among the 20 journals cited by the *Journal of Inorganic Materials,* the only Chinese journal cited was the journal itself. However, the journal that was mostly cited (242 times) was the *Journal of the American Ceramic Society*.

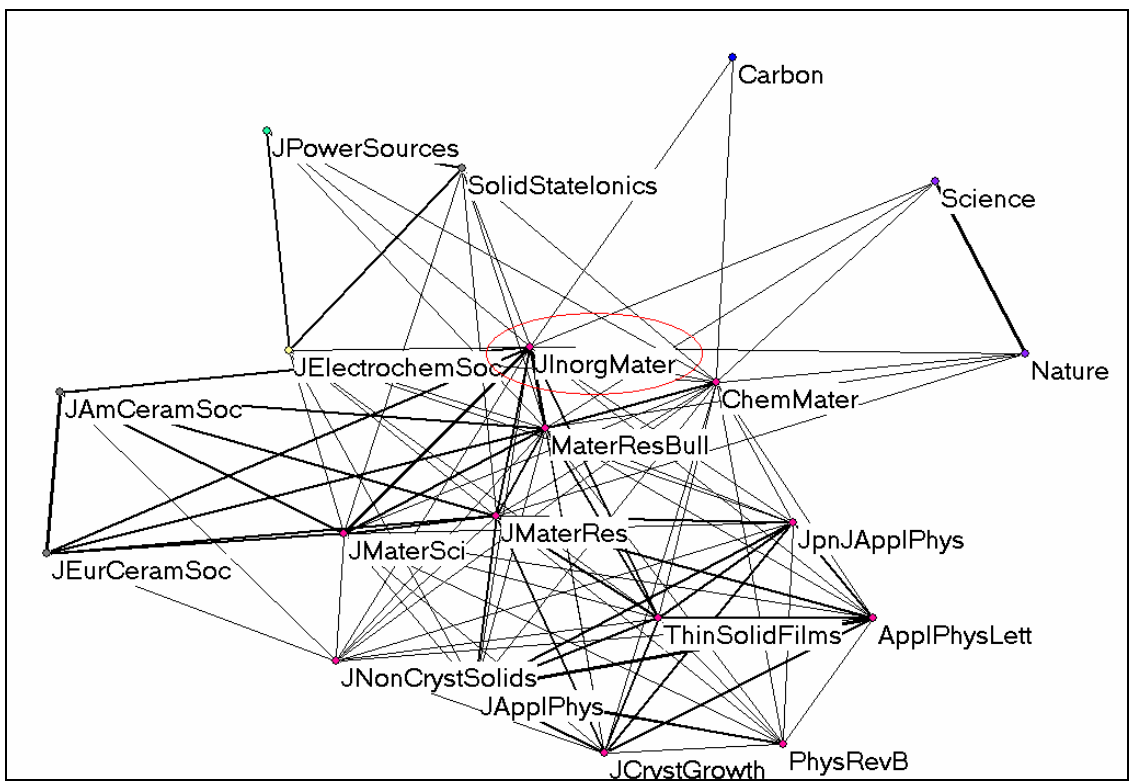

**Figure 5:** Citing environment of the *Journal of Inorganic Materials* (*SCI*, 2003; threshold = 1%; cosine ≥ 0.2).

b.2 Cited pattern

In 2003, the total number of citations of the *Journal of Inorganic Materials* was 346. Thirty-nine journals included in the *SCI* cited the journal two or more times, providing 259 citations to the journal. This is 75% of its total number of citations. The Chinese journals accounted for 56% of this share, and international journals for 19% (Table 2). Of the 56% citation contribution of Chinese journals, 8% were provided by seven English editions and 48% were provided by eleven Chinese editions. Therefore, the international visibility of *JIM* is mainly among Chinese journals, and the journals published in Chinese provided the major citation contribution to *JIM*. When the threshold is set at 1%, 15 journals are included in the cited environment of *JIM*, but only four were international journals; the other eleven journals were Chinese (Figure 6).

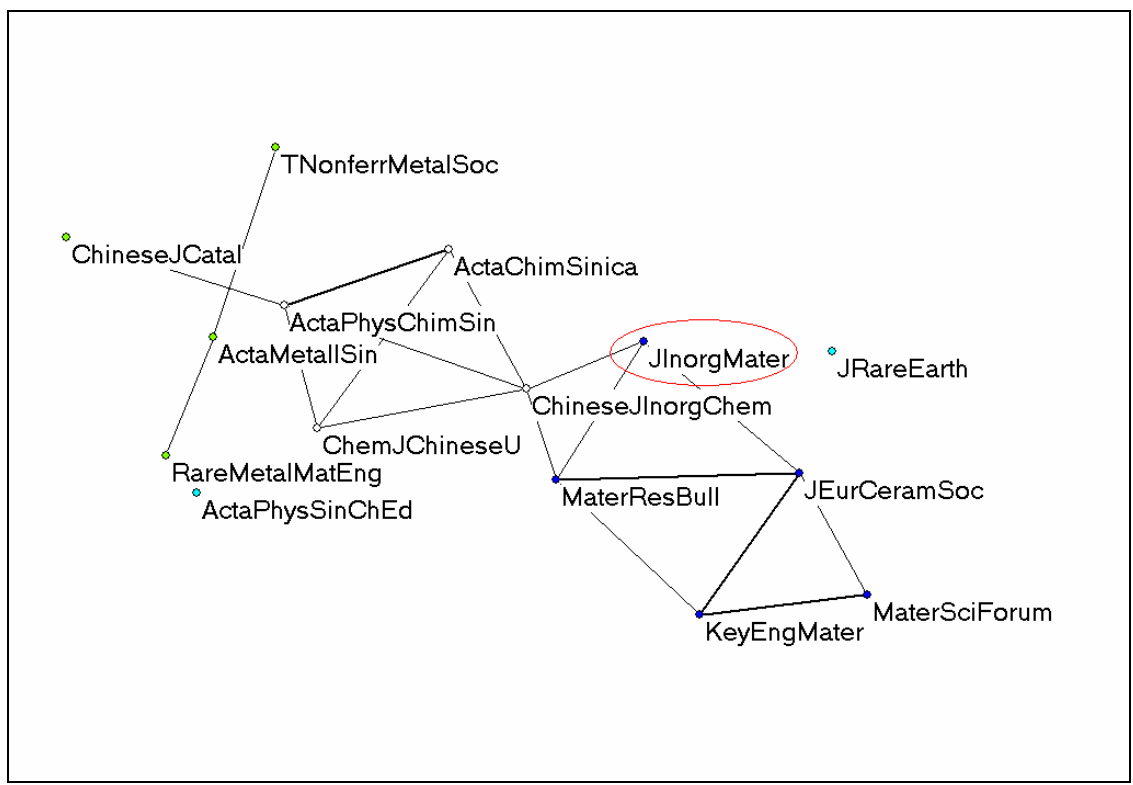

**Figure 6:** Cited environment of the *Journal of Inorganic Materials* in *SCI* 2003 (threshold = 1%; cosine ≥ 0.2).

The above analysis shows that authors in the *Journal of Inorganic Materials* prefer to cite papers in international journals. However, the journal under study does by far not obtain an equal return from its international counterparts.

c. Cited environment of Chinese journals published in English in the *SCI*

Since the *Journal of Inorganic Materials* is published in Chinese, the language barrier may block international scientists from becoming acquainted with its content. Therefore, we chose another journal that is also in material science, but published in English: the *Journal of Materials Science & Technology* (*JMST*), in order to assess whether language indeed functions as an obstacle.

*JMST* was cited 318 times in 2003, among which 211 citations were provided by 43 journals included in the *SCI* and citing *JMST* two or more times. These journals accounted for 67% of the total number of citations, among which 21% was from six Chinese journals and 46% was from international journals (Table 2). Of the 21% share contributed by the six Chinese journals, 12% was from four English-edition journals, and 9% was from two Chinese-edition journals. The international share of number of citations of *JMST* was substantially higher than that of *JIM* (19%).

In 2003, there were 23 journals in the cited environment of the *Journal of Materials Science & Technology* in the *SCI* when the threshold is set at 1%, comprised of 18 international journals and five domestic ones. In addition to higher visibilities among international journals, *JMST* is also integrated with international journals in the graph (Figure 7).

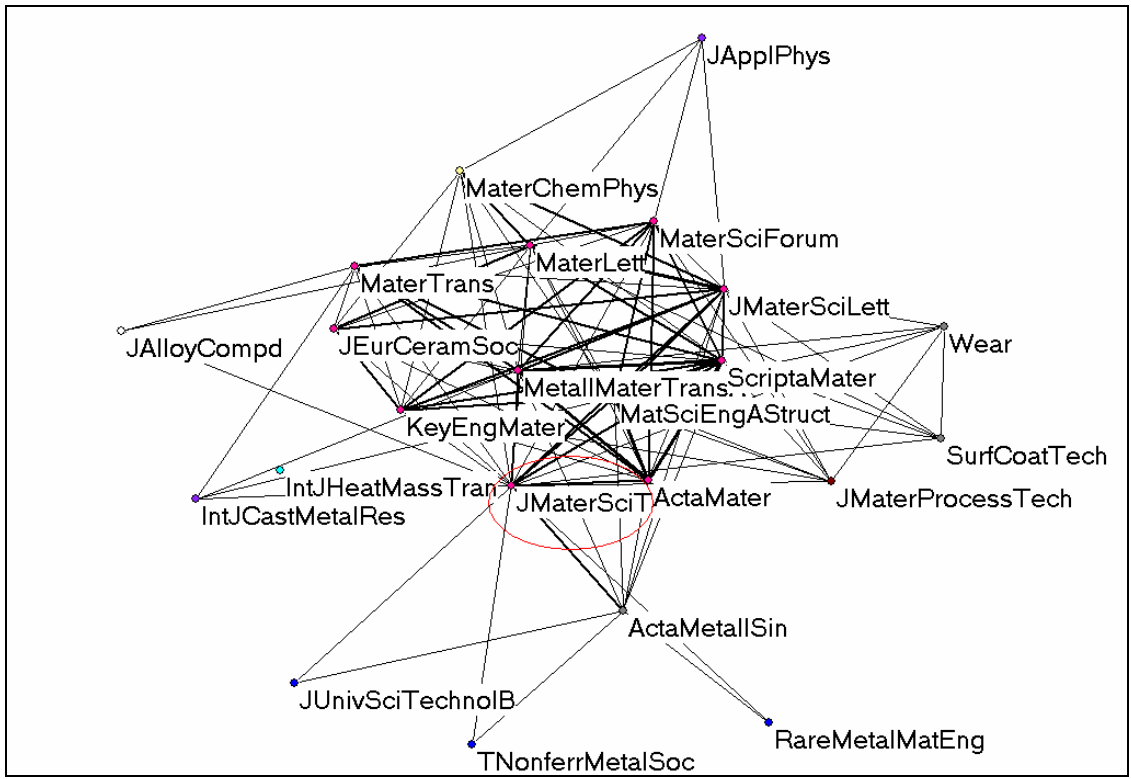

**Figure 7:** Cited environment of the *Journal of Materials Science & Technology* (*SCI* 2003; threshold = 1%; cosine ≥ 0.2).

Whether they are published in Chinese or English, Chinese journals in material science have some international visibility. However, the language factor does affect the connectivity of Chinese journals in the disciplinary network. The international visibility of the two journals in material science demonstrates the difference: the shares of citations from international journals were 19% and 46%, respectively, for the *Journal of Materials Science* (in Chinese) and the *Journal of Material Science & Technology* (in English). Furthermore, the journal in English is incorporated in the graph of its international counterparts, while the other is not.

*3.2.3 Journals in the life sciences*

As mentioned above, China's performance in the life sciences is assessed to be of lower quality when compared to other fields of science, such as mathematics and material science (DICCAS, 2004). We chose a journal that is covered by both the *CSTPCD* and the *SCI* in order to compare the citations or references of journals in this field: the *Science in China Series C-Life Sciences*. This journal has two independent publication editions: the Chinese edition is included in the *CSTPCD* and its English edition is covered by the *SCI*. In this case, the English edition is a duplication of the Chinese one; the *CSTPCD* only includes the Chinese edition, while the *SCI* covers the English edition (Ren & Rousseau, 2004).

a. Citation performance of the *Science in China Series C-Life Sciences* in *CSTPC*

a.1 Citing pattern

The *Science in China Series C-Life Sciences*—we shall indicate this journal with *SCSC-C* for the Chinese edition below—had 1,641 references in 2003, but only 63 journals in the *CSTPCD* received 157 references from this total (10%). This means that 90% of the references of *SCSC-C* were given to journals not included in the *CSTPCD*. As the *CSTPCD* has already covered most of the important Chinese S&T journals, we conjecture again that the other 90% of references are attributed to international journals. Of the 10% (or 157) Chinese references, 23 were given to the *Chinese Science Bulletin*, and 17 to *SCSC-C* itself. Other journals received less than 1% share of the total citations.

a.2 Cited pattern

Although authors in *SCSC-C* have little interest in citing domestic journals, the journal's visibility in the domestic community is high. The journal was cited by 135 journals included in the *CSTPCD* for a total of 282 times, and 25 of these journals satisfied the condition that each contributes to the citations more than 1% (Figure 8). *SCSC-C* gave its highest number of references (23) to the *Chinese Science Bulletin*, and the *Chinese Science Bulletin* gave the same return to *SCSC-C* by contributing the highest number of citations (27) to *SCSC-C*. Figure 8 shows that most of these 25 journals are from the life sciences.

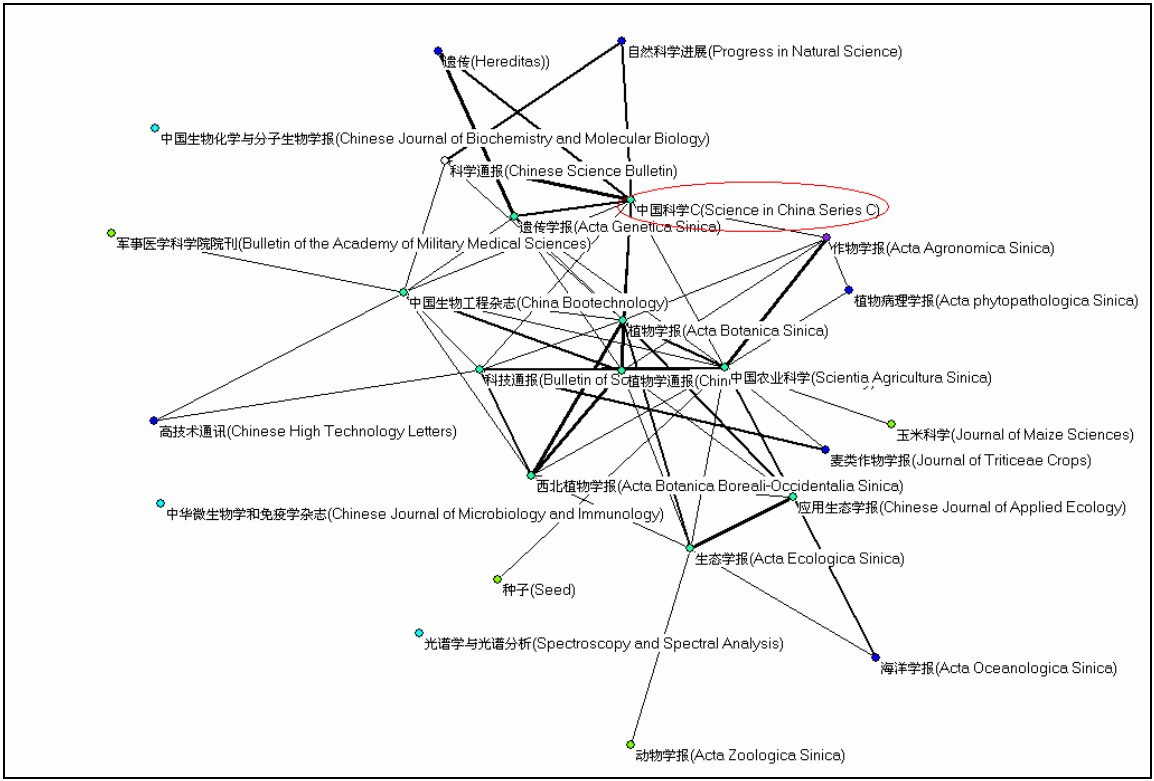

**Figure 8:** Cited environment of *Science in China Series C-Life Sciences* (*CSTPCD* 2003; threshold = 1%; cosine ≥ 0.2).

b. Citation pattern of the *Science in China Series C-Life Sciences* in the *SCI*

b.1 Citing pattern

The English edition of the *Science in China Series C-Life Sciences*—we shall indicate it as *SCSC-E* below—had a total of 1,522 references in 2003; 168 journals in the *SCI* were cited by *SCSC-E* for two or more times. These 168 journals accounted for 1,115 references (73%). Fourteen journals satisfy the condition that each of them provides at least 1% of the total citations of *SCSC-E* (Figure 9).

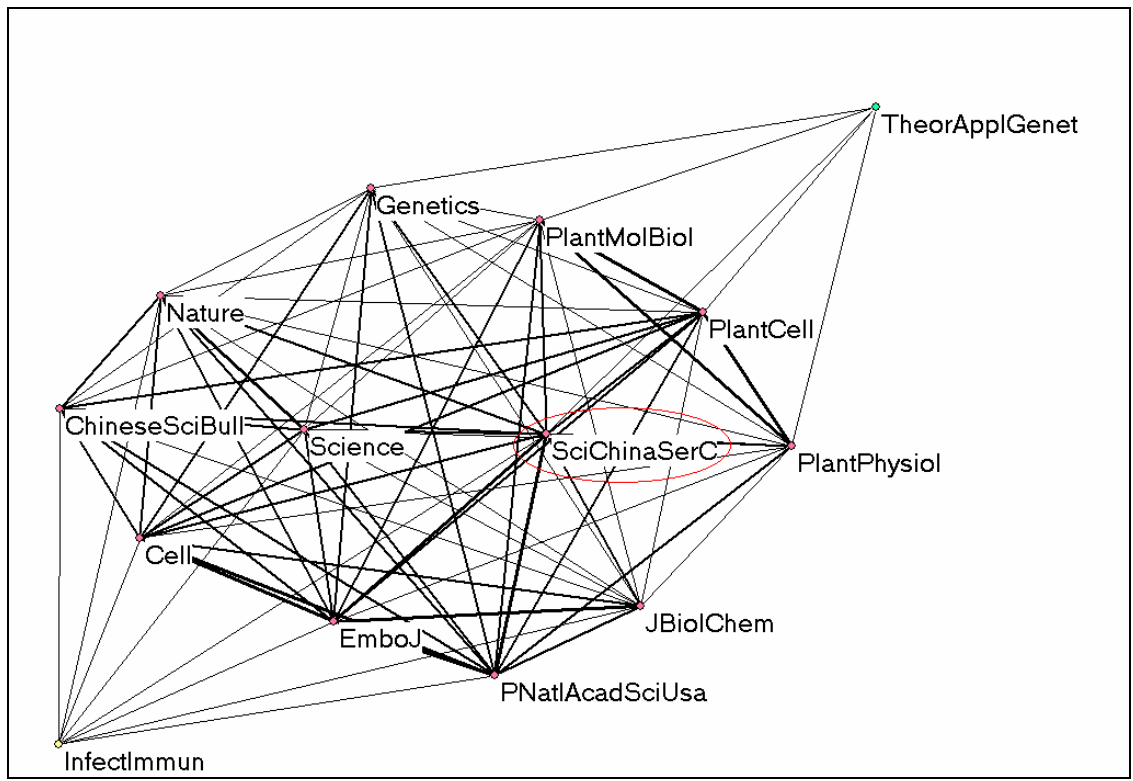

**Figure 9:** Citing environment of *Science in China Series C-Life Sciences* (*SCI* 2003; threshold = 1%; cosine ≥ 0.2).

Journals that achieved the first four highest numbers of references from *SCSC-E* include the *Journal of Biological Chemistry*, the *Proceeding of the National Academy of Sciences of the United States of America*, *Science*, and *Nature*. Except the *Chinese Science Bulletin* and *SCSC-E* itself, the other 12 journals which are heavily cited by *SCSC-E* are international ones. This shows that authors in *SCSC-E* prefer to cite papers published in international journals instead of making references to articles published in domestic journals.

b.2 Cited pattern

Among the number of citations provided by journals that cited the *SCSC-E* at least twice in 2003, 38% were from Chinese journals and 30% were from international journals (Table 2). When the threshold was set at 1%, 15 journals were included in the cited environment of *SCSC-E*. Of these 15 journals, seven are international. Most of the journals citing *SCSC-E* are from the life sciences (Figure 10).

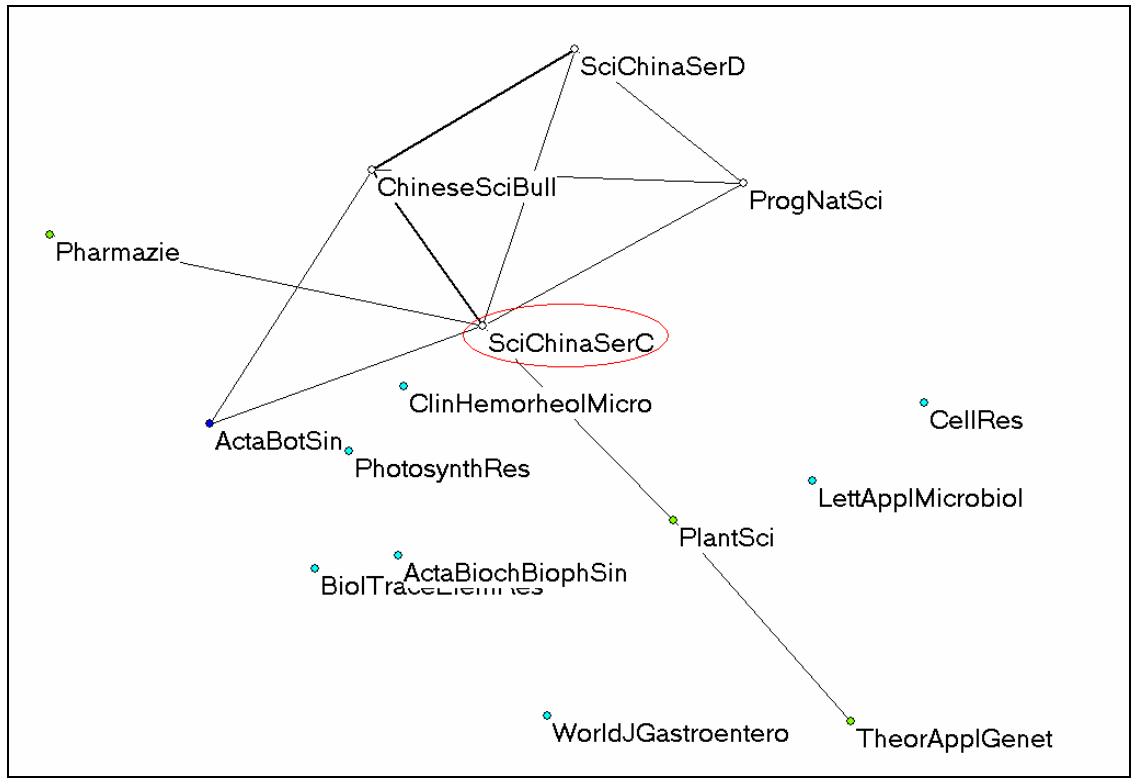

**Figure 10:** Cited environment of the *Science in China Series C-Life Science* (English edition; *SCI* 2003; threshold = 1%; cosine ≥ 0.2).

The citation pattern of the *Science in China Series C-Life Science* shows that this journal's authors, when choosing citations, favour international journals instead of domestic ones. The journal has high domestic visibility, although its international visibility is low. Compared to journals in material science, the international visibility of journals in the life sciences is lower than that of journals in the material sciences (Table 2).

**3.3 Institutional journals**

Among the Chinese S & T journals, more than half (63%) are institutional ones. They are either based in Chinese universities or in the Chinese Academy of Sciences (CAS). The percentage share of university-based journals is 45%, and that of the CAS is 18% (Ren, 2005). In this section we compare the citation patterns of journals from these two types of institutions.

In order to maintain the comparative nature of the study, journals were selected using the following criteria: 1) the journals are from a university and CAS, respectively; 2) they are in the same or similar fields; and 3) they are to be included in both the *CSTPCD* and the *SCI*. Only three Chinese university journals were covered by the *SCI* in 2003; using the criterion that the university journal and the CAS journal should be in the same or at least similar fields, we found two journals that satisfy these conditions: (1) the *Journal of the University of Science and Technology Beijing* and (2) *Science in China Series E-Technological Sciences*. The latter journal is issued by the Chinese Academy of Sciences. Both journals have Chinese and English editions of which the Chinese editions are included in the *CSTPCD* and the English editions are covered by the *SCI*.

Both journals belong to the multidisciplinary category within the classification system provided by the ISTIC staff, and are classified as engineering. The two journals have two independent language editions: one in Chinese and another in English. We studied their domestic visibility through the Chinese editions, and the international relations through the citation patterns of their English editions in the *SCI*. This design enables us to compare their citation patterns and visibilities at home and abroad, respectively.

*3.3.1 Performance in the domestic environment*

In 2003, the *Journal of the University of Science and Technology Beijing* (Chinese edition; *JUSTB-C*) was cited 311 times by 15 journals included in the *CSTPCD*. Most of the reference-providing journals were in material science, or more specifically, metallurgy and metallurgical engineering. In other words, the *Journal of the University of Science and Technology Beijing* is mainly cited by journals in metallurgy and metallurgical engineering (Figure 11).

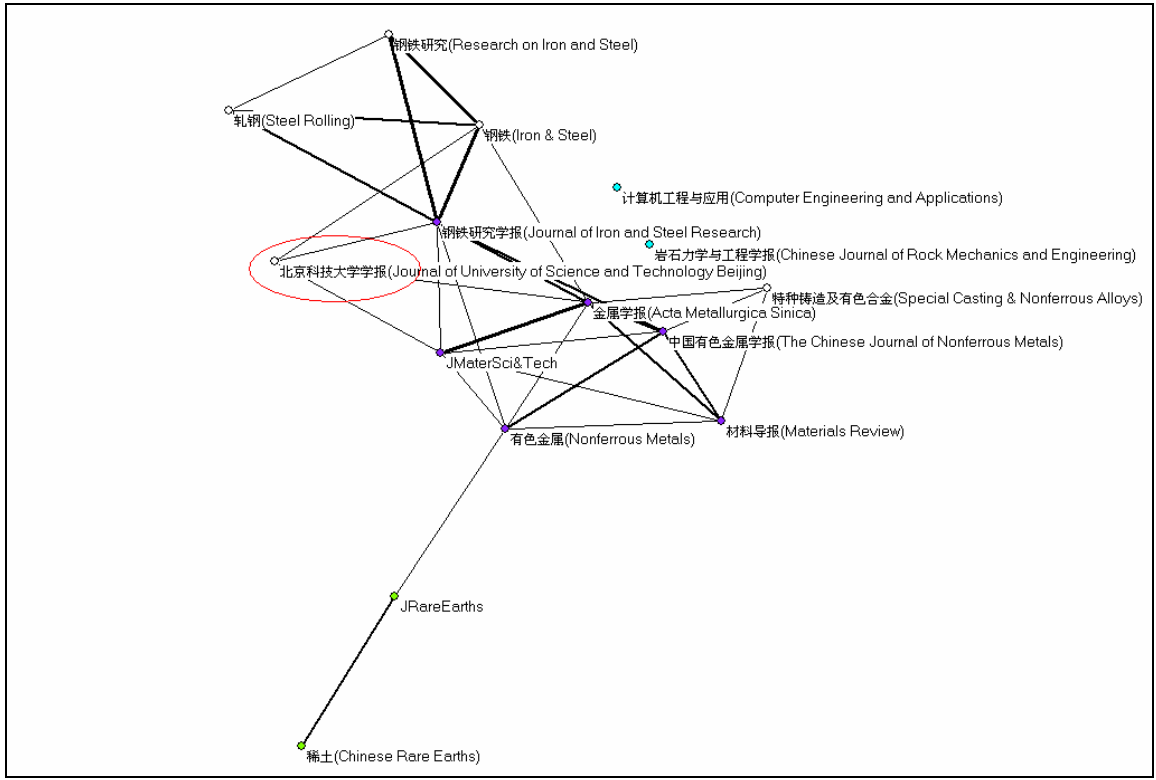

**Figure 11:** Cited pattern of the *Journal of University of Science and Technology Beijing* (Chinese edition, *CSTPCD* 2003; threshold = 1%; cosine ≥ 0.2).

The *Science in China Series E* (Chinese edition, *SCSE-C*) was cited 362 times by 23 journals in the *CSTPCD*. There were eight more journals in the cited environment compared to the *Journal of the University of Science and Technology Beijing* (Chinese edition; *JUSTB-C*). However, the journals citing the *SCSE-C* are mainly in computer science and engineering (Figure 12). Journals in other disciplines like general science, chemistry, physics, mathematics, etc. were also present in the cited environment. Therefore, the *SCSE-C* behaves and is cited more as a multidisciplinary journal with a focus on computing and engineering. Both of the two journals have visibility among domestic journals.

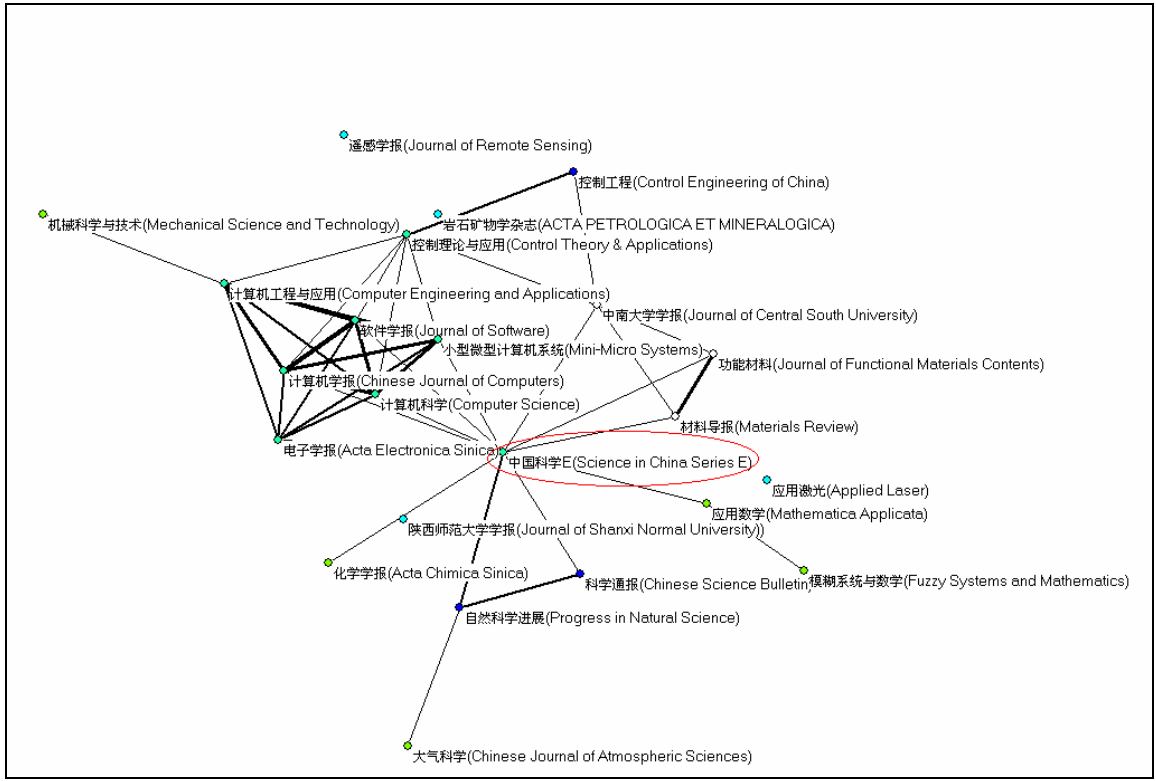

**Figure 12:** Cited pattern of the *Science in China Series E* (*CSTPCD* 2003; threshold = 1%; cosine ≥ 0.2).

*3.3.2 International visibility*

*a.* The *Journal of the University of Science and Technology Beijing*

The cited environment of the *Journal of the University of Science and Technology Beijing* (English edition; *JUSTB-E*) in the *SCI* contained 13 journals when the threshold is set at 1%, among which five were international and eight are Chinese journals. The journal had a total number of citations of 145, among which 64 were within-journal citation (44%). International journals covered by the *SCI* contributed to 11% of its total, while Chinese journals in the *SCI* contributed 64% (Table 2). Thus, Chinese journals made the major contribution in citing *JUSTB-E*. This data shows that the citation impact of *JUSTB-E* mainly happened among Chinese journals. Furthermore, the journal contained a high rate of within-journal citations and it obtained low international visibility.

b. The *Science in China Series E*

There are 20 journals included in the cited environment of the *Science in China Series E* (English edition; *SCSE-E*); among them are eight international journals and 12 domestic ones when the threshold is set at 1%. The journal had a total number of citations of 210, among which only 19 (9%) were within-journal citations. This figure is much less than that of the *Journal of the University of Science and Technology Beijing* (44%). Of the 210 citations, 60% was provided by journals covered by the *SCI*, among which 38% was from Chinese journals and 22% from international ones.

Compared with the *Journal of the University of Science and Technology Beijing*, the *Science in China Series E* performs better in terms of citations by other journals in the *SCI*, since the *Science in China Series E* had less within-journal citations and a higher international visibility.

**4. Conclusions**

Comparison between the two databases (the *SCI* and the *CSTPCD*) shows that except for a fewer number of source journals, the average number of citations per journal in the *CSTPCD* is much less than that of the *SCI*. Chinese authors publishing papers in journals in the *CSTPCD* make less reference to articles in journals than authors publishing papers in journals covered by the *SCI*.

High-quality international journals have a higher elevated rank in the hierarchy than their Chinese counterparts. Authors who publish in high-quality Chinese journals prefer to cite articles in international journals instead of domestic ones. In the case of some high-quality journals, no domestic journals are included in the citation graph when the threshold is set at 1% of their citing environments. However, this tendency does not affect the domestic visibility of these journals. The domestic visibility of high-quality journals in terms of citations is high.

Although authors in Chinese journals prefer to cite papers in international journals above domestic ones, their international counterparts do not provide the same return: the international visibility of Chinese journals is low (Ren & Rousseau, 2002; Liu, M., 1993). The international visibility of Chinese journals differs among disciplines. Among journals in general science, material science and the life sciences, journals in material science have a relatively higher visibility, while journals in the life sciences have the lowest visibility. This reflects the relative strength of China in these fields.

Language is an important factor that affects a journal's visibility. Both Chinese university journals and CAS journals have high domestic visibilities. The international visibility of the CAS journals is higher than that of university journals. Journals in the general sciences are supposed to entertain citation relations with journals in a range of fields. The English edition of the *Chinese Science Bulletin* has this characteristic. The Chinese edition of *Chinese Science Bulletin*, however, focuses more on the geo-sciences in terms of its citation patterns.

Among the journals included in the *SCI* and studied in this paper, the English edition of the *Journal of the University of Science and Technology Beijing* has the highest self-citation rate (44.1%); the self-citation rate of the *Journal of Materials Science & Technology* was the lowest (7.9%). Of the journals included in the CSTPCD and studied in this paper, the Chinese edition of the *Journal of the University of Science and Technology Beijing* has also the highest self-citation rate, while *Science in China Series C* (Chinese edition) has the lowest one (6.0%) (Table 3). These figures show that the visibility of the university journal is low in both the domestic or international scientific communities. The Chinese edition of *Science in China Series C* has high visibility among domestic journals. Journals in material science have the highest international visibility.

| Journal abbreviation | Data source | Total references | Total citations | Self-citations | Self-citing rate (%) | Self-cited rate (%) |
|---|---|---|---|---|---|---|
| *JUSTB-E* | *SCI* | 1338 | 145 | 64 | 4.8 | 44.1 |
| *JIM* | *SCI* | 2788 | 346 | 90 | 3.2 | 26 |
| *JUSTB-C* | *CSTPCD* | 1300 | 311 | 77 | 5.9 | 24.8 |
| *SCSE-C* | *CSTPCD* | 1922 | 136 | 31 | 1.6 | 22.8 |
| *CSB-E* | *SCI* | 12082 | 2302 | 407 | 3.4 | 17.7 |
| *JIM* | *CSTPCD* | 3279 | 896 | 93 | 2.8 | 10.4 |
| *SCSC-E* | *SCI* | 1522 | 228 | 21 | 1.4 | 9.2 |
| *SCSE-E* | *SCI* | 1003 | 210 | 19 | 1.9 | 9 |
| *CSB-C* | *CSTPCD* | 11506 | 3958 | 332 | 2.9 | 8.4 |
| *JMST* | *SCI* | 2656 | 318 | 25 | 0.9 | 7.9 |
| *SCSC-C* | *CSTPCD* | 1641 | 282 | 17 | 1 | 6 |

**Table 3.** Self-citation rates of the journals under study.

**5. Policy implications**

The low visibility of Chinese journals affects the expected number of citations of the papers published within them. As Chinese journals are important channels for Chinese scientists to publish their research results, increasing the visibility of Chinese journals may help to raise the impact of Chinese papers. With the wide influence of the *SCI* in the evaluation of scientific output, inclusion in the *SCI* has already become a major objective among editorial boards of Chinese journals. However, our analysis shows that inclusion in the *SCI* does not necessarily lead to an increase in visibility. More needs to be done to increase this visibility, especially in terms of efforts from both scientific authors and editorial boards.

*5.1 Implications relevant to Chinese authors*

Many factors may affect a journal's visibility. Among them, the intrinsic quality of the papers may play eventually a key role. A highly cited paper is usually creative, original, and makes unique contributions to the relevant fields. The poor citation performance of Chinese articles is one of the important causes of the low visibility of Chinese journals. Among reasons that might lead the low citation performance of Chinese articles are low journal access, poor research quality, emphasis on narrow applications, and selection of research areas outside the mainstream of the communication. The Chinese scientific community has already noticed the low citation performance of Chinese articles and one has made some efforts to change the situation (ISTIC, 2005).

When an excellent piece of research comes to an end, the ability to organize a paper in proficient English becomes a very important factor, since English is a major language in international scientific communication. In general, a journal's visibility relies on the authors that are publishing papers within it. Authors need to enhance not only their academic competence, but also their ability to organize papers in English (Ren, 2004).

The analysis at the level of databases shows that Chinese journals have fewer citations (either number of references or number of citations) per article than that of international ones. This may have the following implications:

- ➢ Chinese authors seem to read less literature than their international counterparts; this may cause Chinese authors to know less about what is occurring in their relevant fields. Being well-informed is helpful for research. If one does not know or knows little about the evolution of his/her research interests, one might conduct research that has already done or partly been done by others, resulting in a misuse of time and money.

- Compared to their international counterparts, Chinese authors have less access to international or even domestic journals, which results in fewer chances to read relevant papers. Among the reasons that can cause such a situation, capital shortage is critical for an institution when one has to decide how much and to which literature one subscribes. Compared to the Chinese rates, prices of international journals are very high and only a few Chinese research institutions or libraries can afford these charges. Even those institutions that can afford a subscription to international journals may have to make choices among interesting literature because of financial shortages.

  Many researchers in China complain that the number of international journals that their institutions subscribe to is too small (Ning, 2002). Let us take the subscription to the *SCI* as an example. Among the 1,396 regular higher education institutions in China, 41 (2.9%) purchased the *SCI* database in 2004 (according to the *SCI* office in China). The accessibility of journals published in China is better than that of international ones; however, many institutions are still puzzled by a shortage of funding. There are three national journal databases providing online services relevant to Chinese scientific publications, but all of them supply service only to users who pay subscription rates. When an institution is not a subscribed user of such web servers, researchers in the institution will face access problems.

- Financial problems also puzzle the editorial boards of Chinese journals. In order to publish more papers within a limited number of pages, editorial boards require authors to limit the number of cited references. Furthermore, authors are forced to cut references in their papers so as to publish more content.

*5.2 Suggestions to the editorial boards of Chinese journals*

With Chinese universities and research institutes encouraging scientists to publish papers in international journals, especially in journals covered by the *SCI*, Chinese journals face fiercer competition in absorbing papers of sufficient quality from Chinese authors (Jin & Rousseau, 2004). On the one hand, Chinese journals stand in a disadvantageous position when competing with their international counterparts because of journal quality and international visibility; on the other hand, such an unfavourable situation forces Chinese journals to improve or reform for survival. This can also be considered as an opportunity for editorial boards to raise the quality of their journals.

The Chinese government has already made a decision to provide some Chinese journals with financial support to help them increase their international visibility (Jia, 2004). But financial aid is not enough to raise a journal's visibility, especially if such financial support reaches only a small number of journals. More efforts need to be made by editorial boards. In addition to absorbing high-quality articles, the following measures might be helpful in improving a journal's visibility:

- ➢ Increasing accessibility for international readers. Journal papers need to be readable before they can be cited. When a journal is easily accessible, the possibility of being cited will increase. To realize this target, Chinese journals need to provide electronic editions, so as to make the content easily accessible through the Internet or specific portals. Until now, there are three national journal databases providing online services relevant to scientific journal publications, and all of these operate commercially (Ren, 2005). However, only one of the three databases—the China National Knowledge Infrastructure, ([www.global.cnki.net/](www.global.cnki.net/))—provides services in English. The other two do not provide English versions and therefore limit their target users to Chinese readers. For the English database website, more promotional work is needed so as to make international scientific readers aware of its availability.

- ➢ Open access. Even though the China National Knowledge Infrastructure provides an English service, its commercial mechanism may prevent international readers from accessing its data, since most international scientists are not paying users. Open access may eliminate this barrier in terms of accessibility.

- Publication of an English edition of the journal. Even though international readers can access Chinese journals, the use of the Chinese language may prevent them from understanding the content of the papers. Using the Chinese language does affect a paper's visibility, and provision of an English version helps to improve visibility.

- Cooperation with international publishers and online journal database providers. International publishers have deliberate strategies for promoting their journals, while online journal providers give direct access to academic researchers. If a Chinese journal is covered by an international journal database, the chance of being read by international researchers will be considerably increased. If made free of charge, universities in other countries may be eager to add Chinese journal collections to their open access databases.

**Acknowledgement**

We are grateful to the statistics team of the Institute of Scientific and Technical Information of China (ISTIC) for supplying relevant data, especially Ma Zheng for providing us with the aggregated journal-journal citation data according to the specification. Wu Yishan offered suggestions for the implications relevant to Chinese scientific authors. We also thank the anonymous referees for their suggestions.

**References:**

Ahlgren, P., Jarneving, B., & Rousseau, R. (2003). Requirement for a Cocitation Similarity Measure, with Special Reference to Pearson's Correlation Coefficient. Journal of the American Society for Information Science and Technology, 54(6), 550-560.

Bensman, S. J. (2001). Bradford's Law and fuzzy sets: Statistical implications for library analyses. IFLA Journal, 27, 238-246.

DICCAS (2004). China in World Science. Beijing: Library of the Chinese Academy of Sciences.

Doreian, P., & Fararo, T. J. (1985). Structural Equivalence in a Journal Network. Journal of the American Society of Information Science, 36, 28-37.

ISTIC, 1998, 中国科技论文统计与分析 (Chinese S&T Papers Statistics and Analysis.) Beijing: Institute of Scientific and Technical Information of China.

ISTIC, 2004. 中国科技期刊引证报告(Chinese S&T Journal Citation Reports 2003). Beijing: Institute of Scientific and Technical Information of China.

ISTIC, 2005. 2004年度中国科技论文统计结果(Statistics of Chinese publications in 2004). Beijing: Institute of Scientific and Technical Information of China.

Jia, H.P. (2004). China gives national science journals a financial boost. Available at http://www.*SCI*dev.net/News/index.cfm?fuseaction=readNews&itemid=1610&language=1

Jin, B.H., & Rousseau, R. (2004). Evaluation of research performance and scientometric indicators in China. In : Moed, H. F., Glanzel, W., & Schmoch, U. (eds) Handbook of quantitative science and technology research. Dordrecht : Kluwer Academic, pp. 497-514.

Jin, B.H., & Wang, B. (1999). Chinese Science Citation Database: its construction and application. Scientometrics, 45, 325-332.

Jones, W. P., & Furnas, G. W. (1987). Pictures of Relevance: A Geometric Analysis of Similarity Measures. Journal of the American Society for Information Science, 36(6), 420-442.

Leydesdorff, L. (1986). The Development of Frames of References. Scientometrics, 9, 103-125.

Leydesdorff, L. (2004). Clusters and Maps of Science Journals Based on Bi-Connected Graphs in the Journal Citation Reports. Journal of Documentation, 60(4), 371-427.

Leydesdorff, L., & Amsterdamska, O. (1990). Dimensions of Citation Analysis. Science, Technology and Human Values, 15 (1990) 305-335.

Leydesdorff, L., & Cozzens, S. E. (1993). The Delineation of Specialties in Terms of Journals Using the Dynamic Journal Set of the Science Citation Index. Scientometrics, 26, 133-154.

Leydesdorff, L., Jin, B.H. (2005). Mapping the Chinese Science Citation Database in terms of aggregated journal-journal citation relations, Journal of the American Society of Information Science & Technology, 56(14) 1469-1479.

Liang, LM (2003). Evaluating China's research performance: How do SCI and Chinese indexes compare? Interdisciplinary Science Reviews, 28(1), 38-43.

Liang, LM, Wu, YS & Li, J. (2001). Selection of databases, indicators and models for evaluating research performance of Chinese universities, Research Evaluation, 10(2), 105-113.

Liu, M. (1993). A study of citing motivation of Chinese scientists, Journal of Information Science 19, 13-23.

Ministry of Education (2003). 年全国普通高等学校人文、社会科学人力情况. Beijing: Ministry of Education. Available at
http://www.moe.edu.cn/edoas/website18/info15069.htm

Narin, F., Carpenter, M. & Berlt, N. C. (1972). Interrelationships of Scientific Journals. Journal of the American Society for Information Science, 23, 323-331.

Ning, W. (2002), 科技会议大家谈(Forum on Science and Technology). Available at:
http://202.114.201.91/oldxb/files6/2002/0101.htm

Ren, S.L. (2004). 英语科技论文撰写与投稿 (How to write and publish scientific papers in English). Beijing: Science Press.

Ren, S.L. (2005). Editing scientific journals in Mainland China, European Science Editing February, 31(1). 8-9.

Ren, S.L. & Rousseau, R. (2002). International visibility of Chinese scientific journals, Scientometrics, 53(3), 389-405.

Ren, S.L. & Rousseau, R. (2004). The role of China's English-language scientific journals in scientific communication, Learned Publishing, 17, 99-104.

Seglen, P. O. (1997). Why the impact factor of journals should not be used for evaluating research, British Medical Journal, 314:497.

Salton, G., & McGill, M. J. (1983). Introduction to Modern Information Retrieval. Auckland, etc.: McGraw-Hill.

Simon, H. A. (1973). The Organization of Complex Systems. In H. H. Pattee (Ed.), Hierarchy Theory: The Challenge of Complex Systems (pp. 1-27). New York: George Braziller Inc.

Sivertsen, G. (2003). Bibliografiske Datakilder Til Dokumentasjon Av Vitenskapelige Publikasjoner. Oslo: NIFU skriftserie nr. 22/2003; at http://www.nifustep.no/norsk/publikasjoner/bibliografiske_datakilder_til_dokumentasjon_av_vitenskapelige_publikasjoner.

Tijssen, R., de Leeuw, J. & van Raan, A. F. J. (1987). Quasi-Correspondence Analysis on Square Scientometric Transaction Matrices. Scientometrics 11, 347-361.

Van Leeuwen, T. N., Moed, H. F., Tijssen, R. J.W., Visser, M. S., & Van Raan, A. J. (2001). Language biases in the coverage of the Science Citation Index and its consequences for international comparisons of national research performance. Scientometrics, 51, 335-346.

Wu, Y.S., Pan, Y.T., Zhang, Y.H., Ma, Z., Pang, J.A., Guo, H., Xu, B., & Yang Z.Q. (2004). China scientific and technical papers and citations (CSTPC): history, impact and outlook. Scientometrics, 3, 385-397.

Zhou, P., & Leydesdorff, L. (2006). The emergence of China as a leading nation in science, Research Policy, 35(1), 83-104.